\documentclass[a4paper,10pt]{article}
\usepackage[utf8]{inputenc}
\usepackage{authblk}
\usepackage{graphicx}
\usepackage{placeins}
\usepackage{url}

\title{A gas-phase reaction cell for modern Atom Probe systems}
\author{D. Haley${}^1$, I. McCarrol${}^{2,3}$, P. A. J. Bagot${}^1$, J. M. Cairney${}^{2,3}$, M.P. Moody${}^1$}
\date{}
\affil{\small{${}^1$Department of Materials, Oxford University, 16 Parks Road, Oxford, OX1 3PH, UK\\
${}^2$ School of Aerospace, Mechanical and Mechatronic Engineering, The University of Sydney, NSW 2006, Australia\\
${}^3$Australian Centre for Microscopy and Microanalysis, University of Sydney, Madsen Building F09, NSW 2006, Australia}}

\begin{document}

\maketitle

\begin{abstract}
In this work, we demonstrate a new system for the examination of gas interactions with surfaces via Atom Probe Tomography. This system provides the capability to examine the surface and subsurface interactions of gases with a wide range of specimens, as well as a selection of input gas types. This system has been primarily developed to aid the investigation of hydrogen interactions with metallurgical samples, to better understand the phenomenon of hydrogen embrittlement. In its current form, it is able to operate at pressures from $10^{-6}$ to $1000$~mbar (abs), can operate using a variety of gasses, and is equipped with heating and cryogenic quenching capabilities. We use this system to examine the interaction of hydrogen with Pd, as well as the interaction of water vapour and oxygen in Mg samples.

\end{abstract}

\section{Introduction}
It is difficult to provide quantitative analysis of the behaviour of hydrogen in engineering materials at the nano-scale. Currently, there is no universal characterisation method that can analyse for hydrogen at such small length scales, with simultaneous spatial and chemical accuracy. Hydrogen is however, extremely important, owing to its near-ubiquity in operating environments.  A better understanding of the behaviour and consequences of hydrogen is extremely important from an industrial perspective, for marine systems~\cite{Banerjee1999}, structural materials, bearings in turbines~\cite{Evans2016} and energy storage~\cite{DRIVEPartnership2013}. Despite the extent of this issue, the nature of hydrogen's diffusivity, low-concentration in many metallurgical systems and low scattering cross-sections under X-ray, electron and neutron illumination, mean that many standard techniques are of diminished utility when examining for hydrogen. Futhermore, hydrogen is critical in several deleterious phenomena, such as embrittlement in steels, where sudden and difficult to predict failure can occur~\cite{Archakov1985}\cite{Oriani1987}. The lack of imaging capacity for hydrogen has contributed to the difficulty of understanding such issues, hence new imaging capacity at the nano-scale may aid in understanding such processes in more detail, however the development of new experimental techniques is needed to achieve this goal.

Atom Probe Tomography (APT) is a microscopy technique  enabling the examination of the atomic-scale nature of a material, with ppm-level chemical sensitivity and nanometre spatial accuracy. APT samples are in the form of small needles, wherein ions are evaporated from the apex of these needles and projected onto a two-dimensional detector. From these APT experiments a three-dimensional reconstruction of the original material, prior to its evaporation can be generated. The obtained dataset contains the positions and chemistry of the atoms evaporated from the sample - providing simultaneous ppm-level chemistry with nm-level spatial information. 

The experiment must be undertaken under high-vacuum ($10^{-10}$~mbar) conditions. This ensures that there is minimal adsorped contamination on the cold sample stage, due to reduced atomic arrival rates, and minimises the chance of an ion being scattered en-route to the detector. The evaporation process at the beginning of the experiment has the additional effect of removing any contaminants that may be present on the sample, leaving a highly clean sample surface. The cleaned sample surface presents an ideal target to further probe the interaction of molecular species with surfaces, via the controlled introduction of additional gas species. 

There has been an extensive history of the introduction of gases into the high vacuum chamber itself, such as to aid the Field Ionisation Microscopy (FIM) method~\cite{Brenner1970}. Indeed it is the introduction of an inert imaging gas at low pressures in FIM that enables atom surface positions to be imaged in a reliable manner. Further studies have been undertaken to examine the interaction of non-inert gases with the surfaces of samples~\cite{Bocarme2009}\cite{Bagot2006}, including examination of D${}_2$ at surfaces~\cite{Kruse2001}, as well as for the introduction of deuterium directly into APT samples~\cite{Takahashi2010}. 

In the case of deuterium, this is used to separate out the relative contributions of hydrogen introduced into the sample from the UHV chamber during the experiment, from that in the sample, using an isotopic tracing method. Indeed, researchers have used APT to examine deuterated systems using an ex-situ approach~\cite{Karnesky2012}. Alternately, others have investigated methods to suppress H levels, to aid in the interpretation of mass spectra~\cite{Sundell2013}. However in-situ charging approaches should enable a more direct and controlled method for the exposure of samples to gas environments, providing reduced contamination and the ability to examine systems that may otherwise react with air. 

Here we detail our system for controlled environmental exposure that has been developed for modern atom probe systems. The system is capable of providing controlled gas pressures over a wide pressure range, whilst also permitting UHV transfer between the atom probe and the reaction chamber without air exposure. Heating facilities incorporated into the reaction chamber allow for exposure at up to $400^{\circ}$C, with post-exposure cooling also provided via an $\mathrm{LN}_2$ backed cold-finger. Transfer from the reaction chamber can be performed in minutes, with cryogenic cooling allowing for rapid quenching.

We demonstrate the application of this system for deuterium charging of Pd, to demonstrate the potential for this apparatus to examine the behaviour of hydrogen within these materials. We perform our atom probe analyses under laser illumination, and extend our analysis protocol for detecting hydrogen within APT.\@ We show that in Pd significant levels of deuterium can be taken up, detected in APT and then removed by heating. We further suggest that in our experiments the ${}^1$H signal alone may not be a suitable indicator of hydrogen uptake, and paired charged/uncharged experiments may be necessary to fully investigate H uptake in materials. Lastly, we demonstrate that this system is not limited to the analysis of hydrogen, demonstrating Mg oxidation results that can be used to examine the interaction of oxygen and water vapour with differing alloy surfaces.

\section{Apparatus}
\begin{figure}[h]
 \centering
 \includegraphics[width=0.95 \textwidth]{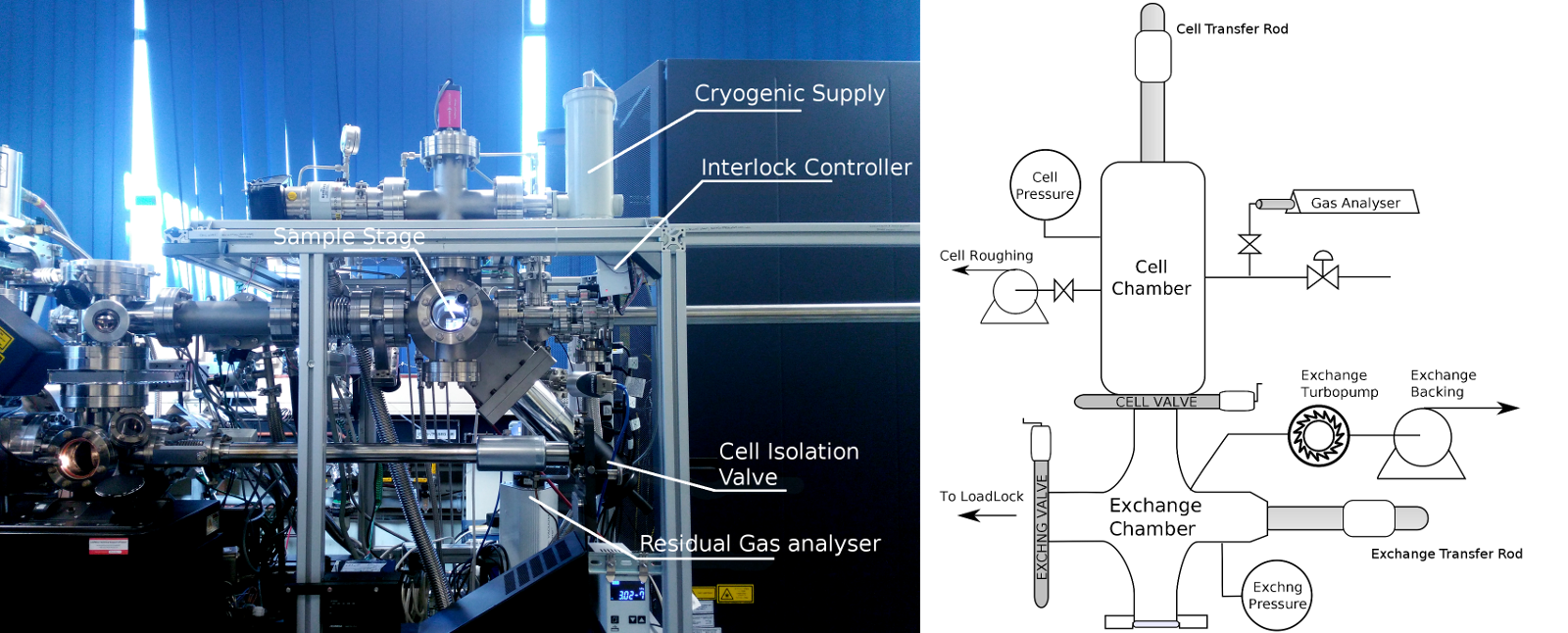}
 % reactionCellPhoto.png: 0x0 pixel, 300dpi, 0.00x0.00 cm, bb=
 \caption{Photograph and diagram of the reaction cell system described in this work, as attached to a LEAP3000X-HR. Cell system consists of a surrounding support structure that allows for interchange of samples into the atom probe load-lock. The system is mostly independent of the operation of the LEAP system, and is detachable.}
 \label{fig:reactionCellPhoto}
\end{figure}

The cell apparatus constructed in this work (Figure~\ref{fig:reactionCellPhoto}) is an additional attachment for a Cameca LEAP3000X-HR atom probe system. The cell forms a stand-alone attachment that can be operated nearly independently of the atom probe. The transfer attachment point is made at the pre-existing turbomolecular pump port located on the atom probe loading port. Anti-tilt attachments for mechanical stability are clamped onto the LEAP frame, however the cell is otherwise a free-standing system that sits above the LEAP transfer arm to minimise floor-space usage. The cell frame houses the cell itself, gas manifolds and UHV control devices. Pumps are located below the cell frame on vibration dampening mounts. 

The cell itself consists of two chambers with suitable pumping equipment to minimise contamination into the LEAP system. The first port is the exchange chamber, which provides a HV pumping environment, and is isolated from the cell via a gate valve. The primary reaction chamber, connected to the exchange chamber, permits the introduction of a gas environment, and can be directly pumped to base pressures at $10^{-2}$~mBar levels. Once evacuated to these levels, the isolation valve is opened and the exchange chamber can be used to further reduce pressures to HV levels, allowing both chambers to be evacuated using a turbomolecular pump. A Residual Gas Analyser (RGA) in the reaction chamber allows for the qualitative monitoring of vacuum quality. This device is isolatable from the cell to prevent contamination issues.

Samples, in their ``pucks'' (sample mounts), can be transferred to and from the APT loading port using a magnetically coupled transfer rod. To prevent any sliding motion of the atom probe's ``carousel'' sample puck holders, the LEAP's vertical transfer rod must be used to support the carousel during sample transfer. 

The gas manifold allows for the introduction and control of gas sources, and is of 316 stainless steel construction. Within the gas manifold, coarse pressures are monitored using a dry Bourdon gauge, from rough vacuum to 5~atm. Roughing pump systems utilise a nitrogen gas ballast to minimise explosion risk. When not in use, the manifold is held under rough vacuum, to mitigate contamination concerns.

The sample stage assembly within the reaction chamber is attached to the end of a magnetic UHV transfer rod. Heating on the stage is provided by an electric resistive cartridge heater, sealed in a steel jacket. A ferritic stainless block is used as a heat spreader to ensure good thermal contact with the sample stage, and to provide a connection to the type K thermocouple used for temperature control input. Control is provided by a PID control box (Dwyer, 16A series), using a solid-state relay using pulse-width modulation to provide smooth control. The stage assembly is thermally isolated from the stage by a thick ceramic spacer. Wiring to and from the stage is heat-sinked against the shaft of the transfer rod, where a kapton coated wire is used to provide electrical insulation. For simplicity, the sample stage itself is a modified benchtop puck holder, and is constructed of an aluminium alloy. However, this limits the maximum operating temperatures to 450$^\circ$C in the current configuration.

\subsection{Software}
The system is controlled using a custom-developed software program, to allow for the control of the attached devices. These are controlled over a serial (RS-232) link which enables software-based operation, logging and interlocking of the system. There are several key devices to control, including Pfeiffer and MVC-3A HV gauge controllers, as well as a Pfeiffer Prisma quadrapole mass analyser (QMA-200), and finally Edwards nXDS-6 pumps. To operate these devices we developed a modular HV device communications library, \emph{libvacuumdevice}. This library is a C++ library that provides connection negotiation and communications, built upon libserialport. Datalogging is performed on all measured gauges, and the system can in future be further adapted to include heater values via MODBUS.\@ The source-code is available at \url{http://apttools.sourceforge.net/}, and can be re-utilised in other UHV applications.

\section{Methods}
\subsection{Cell and atom probe}
Experiments were performed using a LEAP3000X-HR with a 532~nm pulsed laser, equipped with the custom apparatus described above. APT analyses were performed under laser illumination, in order to mitigate yield concerns arising in voltage mode experiments.  In all cases, no uncontrolled air exposure occurred after initial preparation and loading of the sample into the atom probe.

For safety reasons, deuterium was generated using an electrolytic hydrogen generator (SRI), using 99.8\% Deuterium heavy water (Alfa-Aesar) as the water source. To ensure that gas can be introduced quickly into the chamber, a 1~L pressure tank was used to store generated hydrogen, before it was admitted into the chamber. Prior to gas charging, all lines were evacuated through the reaction cell chamber, for several minutes, to minimise contamination from residual gas sources. For deuteration, the stage was pre-heated to 200${}^\circ$C before heating samples, to out-gas any volatile contaminants that may have adsorbed onto the stage, such as $\mathrm{H}_2\mathrm{O}$. The gas loading stage takes $\approx 1$~min to fully load the chamber, and is controlled by means of a manual throttling valve.

Evacuation of the gas is in two stages, in the first stage gas is extracted using a roughing pump. In the second stage, the exchange chamber turbomolecular pump is used to reduce the base pressure of the system back to high vacuum levels. This two stage process takes on the order of 2-3 minutes to return to HV levels, depending on input gas. RGA scans were taken both before and after gas introduction, to monitor for sufficient vacuum recovery. 

Oxidation experiments were carried out using $\mathrm{O_2}$ gas and $\mathrm{H_2O}$ vapour, at room temperature. The $\mathrm{H_2O}$ was produced by passing dry compressed air through distilled water, and a dry-air bypass stream was used to control the final humidity. The gas piping system was purged with $\mathrm{N_2}$ both before and after gas exposure. Exposures details for each sample are given in Table~\ref{tab:exposureData}.

\subsection{Samples}
{%
\newcommand{\mc}[3]{\multicolumn{#1}{#2}{#3}}
\begin{table}
\centering
\caption{Exposure parameters for samples in this reaction system.}

\begin{tabular}{lccl}\hline
 & Time (min) & Pressure (mBar) & \mc{1}{c}{Gas}\\ \hline
 Pd & 15 & 1000 & $\mathrm{D_2}$ \\
Mg40Fe & 150 & 200 & $\mathrm{O}_2$\\
Mg880Ge & 60 & 500 & $\mathrm{O}_2$\\
Mg650Fe & 5 & 700 & $\mathrm{H_2}\mathrm{O}/\mathrm{Air}$, 85\%RH \\ \hline
\end{tabular}
\label{tab:exposureData}
\end{table}

}%
To generate the required needle shaped samples,  wire samples of Pd (Goodfellow, 99.95\%) were etched in 2\%-perchloric acid in 2-butoxyethanol, using a standard electropolishing configuration. Similarly Mg samples were prepared by mechanical cutting (under ethanol) followed by electropolishing, then a ``rough polishing'' step in 10\% perchloric acid in methanol before standard electropolishing. The final needle-shaped samples, crimped into copper tubes, were then loaded into a sample holder (``puck'') composed of either an Al (for heating experiments) or a Cu body both with a stainless steel grip mechanism.

Subsequently, samples were loaded into the atom probe (Mg samples within 20 minutes of polishing), and an initial analysis was undertaken. The Pd samples were transferred to the reaction chamber for heating at 250${}^\circ$C and subsequent re-analysis by APT.\@ This aimed to determine whether any significant pre-loaded quantity of ${}^1$H was present, as well as to determine the levels of contamination that were present. For each analysis, \textgreater500,000 ions were collected, at a set temperature of 55~K, a pulse frequency of 100~kHz (to avoid ``wrap-around'' problems from high-mass elements~\cite{Larson2013}), and 0.2~nJ/pulse laser energy.

Mg samples were run at 0.2,0.4 and 0.5~nJ laser energy and 100, 160 and 200~kHz and 30, 30 and 50~K for the Mg-(40Fe, 880Ge, 650Fe) samples respectively, where e.g. Mg40Fe designates Mg + 40ppmFe.  Variation in these values was due to attempts to optimise yield, whilst maximising mass spectral resolution - the low yield in the samples meant that the search space has to date been only sampled sparsely. These datasets display the versatility of the constructed system and the fact that the sample conditions are different for each reflects this versatility.

After APT analysis post-heating, Pd samples were then exposed to a ${}^2$H atmosphere at room temperature and 1000~mBar (abs) for 15 minutes. After evacuation of the gas, samples were transferred back to the atom probe's UHV environment for subsequent analysis.  When examining mass spectra within a decomposition analysis, significant mass-to-charge overlap concerns arise from the individual ionic contributions of each species, during the analysis of hydrogen charged samples. Subsequently an open-source overlap solving program was employed to solve these complex overlaps~\cite{London2017} (Software revision 7f05f2269efa). This program was modified to include deuterium as its own pseudo-element (an ``element'' called `D' at mass 2, with 100~\% abundance), whilst also retaining the naturally abundant hydrogen.

\section{Results}
\subsection{Palladium}
For the Pd samples, there are three analysed conditions, as prepared, post-heating and post deuterium charging, all performed on the same specimen.  In the as-prepared case $\mathrm{H}_2^+$ to $\mathrm{H}^+$ ratios (prior to any heating or charging) were found to be 0.187, with an apparent H content of 2.5~at\% (0.7~at\% detected as $\mathrm{H^{{}+{}}}$).  After heating, this ratio was found to be somewhat changed, at 0.631, and an apparent H content of 3.1~at\% (1.38~at\% as $\mathrm{H}^+$). Between the two mass spectra, the primary change was an apparent increase in $\mathrm{PdH^+}$ after heating.  There did not appear to be any significant change in charge state ratios or steady-state operating voltage after heating, suggesting the tip geometry was relatively unchanged. 

\begin{figure}[h]
 \centering
 \includegraphics[width=0.75 \textwidth]{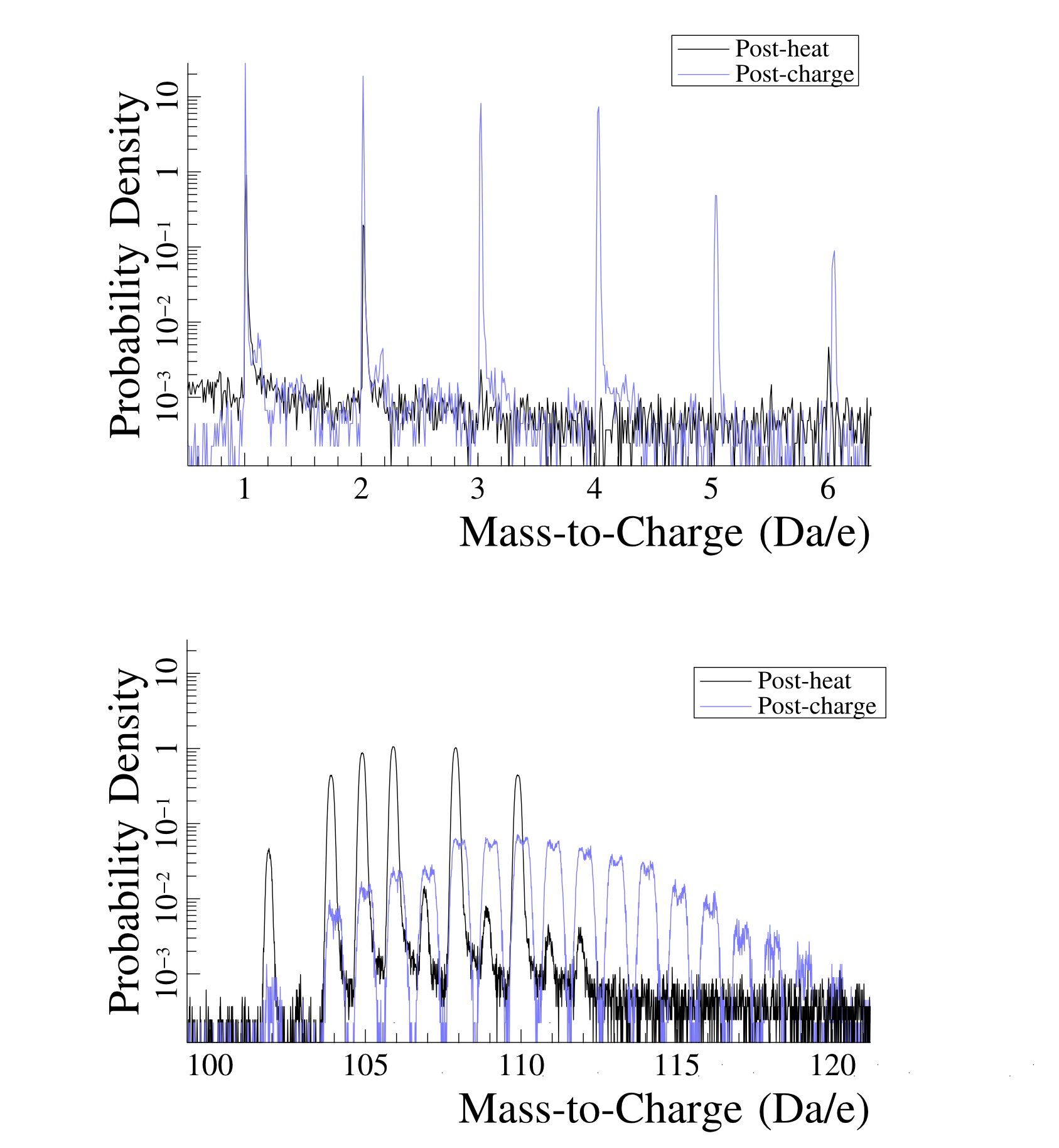}
 % mass_spectrum_pre_post_charge_combined.png: 0x0 pixel, 300dpi, 0.00x0.00 cm, bb=
 \caption{Sub-sections of Pd-mass spectrum prior to and post charging. Sample had been heated prior to charging to ensure all internal hydrogen had been removed. Note marked shift in Pd mass spectrum peaks, indicating uptake of gas due to deuterium charge. ``Splitting'' of charged peaks is due to voltage/bowl corrections to the raw time-of-flight measurement, in software, and are not present in the raw time-of-flight data.}
 \label{fig:pdMassSpectra}
\end{figure}
\begin{figure}[h]
 \centering
 \includegraphics[width=0.8 \textwidth]{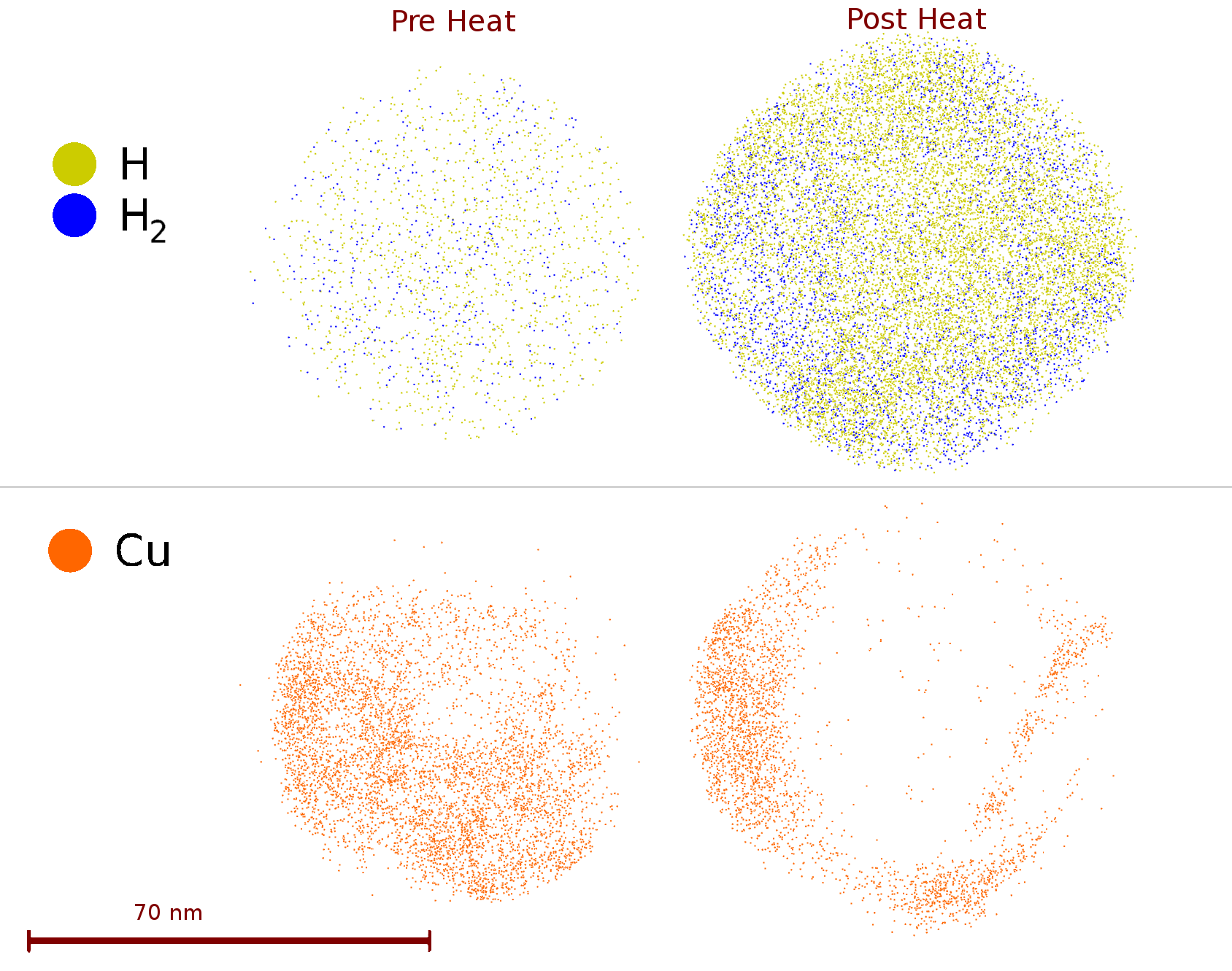}
 % Cu_contamination_and_surface_restructuring.png: 0x0 pixel, 300dpi, 0.00x0.00 cm, bb=
 \caption{Surface restructuring during sample heating (top), leading to differing $\mathrm{H_2}$ and H spatial distributions, attributable to varying electric field intensity across the surface.  Copper contamination from electropolishing (bottom left) and from residual contamination from both polishing and heating of the sample tube (bottom right). Cu is observed in experiments where Cu holders are used.}
 \label{fig:cucontamination}
\end{figure}


fter 3D reconstruction, a small quantity of Cu was found to be deposited onto the sample, both before and after heating, while the Cu is located at an additional position on the sample post-heating. (Figure~\ref{fig:cucontamination}). Some oxygen, as $\mathrm{O}_2^{2+}$/$\mathrm{O}^+$ was found, with clear segregation to the crystallographic poles post-heat (0.32~at\% prior to heating, 0.155~at\% (ionic basis)), not visible previously. This observed oxygen is an artefact arising from the reaction, rather than e.g.\ field migration, as we do not observe it in the prior-to-heating experiments - possibly reactions with oxygen containing contaminants, such as $\mathrm{H}_2\mathrm{O}$ are occurring at specific crystal orientations. The heating tests also show that there is minimal contamination by other sources, such as $\mathrm{H_2O}$, or C. Further experiments utilising SS316 tubes are under-way to minimise such Cu contamination.

\begin{figure}[h]
 \centering
 \includegraphics[width=0.75 \textwidth]{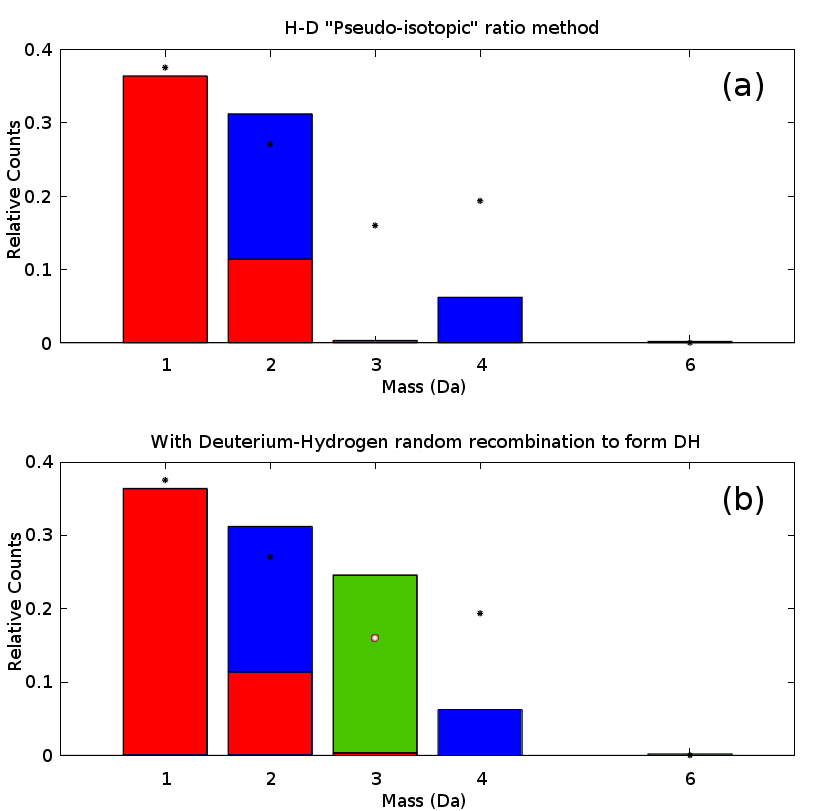}
 % PdD_H_peaks_decomposition.png: 0x0 pixel, 300dpi, 0.00x0.00 cm, bb=
 \caption{Pseudo-isotopic ratio (fixed $\mathrm{H}_2^{+}$:H$^{+}$ ratio) decomposition of hydrogen (${}^1$H${}^+$ and ${}^2$H${}^+$) peaks - (a) without DH and (b) with DH recombination. Markers show measurements, and bars are deconvolved fit (least-squares). Note that the 4~Da peak in both cases is not well explained under this model. The 3~Da peak in (b) is formed by assuming random recombination of D and H at the same rate as the H-$\mathrm{H}_2^+$ recombination, using concentrations from (a). Mass 6 is $\mathrm{D}_3^+$, which is distinguishable from $\mathrm{C}^{2+}$, due to sufficient mass separation.}
 \label{fig:iterativeDecompositionH}
\end{figure}

ollowing deuterium charging of the sample, there is a distinct alteration in the mass spectrum, with peaks visible from mass-to-charge ratios of 1-6~Da  (which correspond to the set of peaks $\mathrm{H^+/H_2^+/D^+/DH^+/D_2^+/D_2H^+/D_3^+}$), and $\mathrm{PdH^+/Pd^+/PdD^+}$ peaks from mass 102-120~Da, shown in Figure~\ref{fig:pdMassSpectra}. The low mass peaks originate as cluster ions, for which the trimer $\mathrm{(H/D)_3^+}$ is stable at low fields~\cite{Miller1996}, and unstable at higher fields. The cross-terms, such as $DH^+$ are due to recombination of introduced deuterium and hydrogen from other sources. The $\mathrm{PdH^+/Pd^+/PdD^+}$ peaks can be explicitly decomposed, as these have sufficient numerical rank to analyse, for a reasonable set of trial species. However, to decompose the H peaks at (mass $\leq$ 6), the problem is rank-deficient (has an infinite number of solutions) when treated as a traditional decomposition problem.  Examining Figure~\ref{fig:pdMassSpectra}, we do not observe significant peaks at non-integral masses, indicating that the species are likely singly charged. When utilising the `weights' program to examine the observed peaks, species that may be of interest include PdN and PdOH.\@ Whilst PdC is a candidate, it did not provide any reasonable contribution when added to the linear decomposition, and was judged to be a poor fit. Whilst this can be solved, further information on suitable trial species would be valuable.

he problem of rank deficiency can be solved, given sufficient assumptions. One possibility is to obtain deuterium-species formation rates are based upon an initial decomposition using a limited selection of isotopes, thereby assuming only H and D combine to form their respective peaks at mass 1 and 2, with $\mathrm{H}_2^+:\mathrm{H}^+$ in the same ratio as the pre-charge experiment (Figure~\ref{fig:iterativeDecompositionH}a) - thus D forms a separate linear basis ($\mathrm{D}^+$ and $\mathrm{D_2}^+$), akin to a pseudo-isotopic ``fingerprint'', fully distinguishable from the fixed $\mathrm{H^+:H_2^+}$ peak ratio. However, simply assuming this fits the observed data rather poorly. However, in addition to this fixed ratio assumption, the cross-term (e.g.\ DH${}^+$) can be computed - by allowing H and D to recombine to form $\mathrm{HD}^+$ (Figure~\ref{fig:iterativeDecompositionH}b). This however retains a significant deviation from the observations at mass 4~Da, indicating either a change in propensity for D$\rightarrow$D${}_2$ recombination, or more disconcertingly, that the prior recombination rate ($\mathrm{H}^+_2$:H${}^+$ ratio) has only limited validity. This will be discussed in more detail in a future work.

nterestingly, there is still a large ``false'' hydrogen peak according to this decomposition, that is not part of the material. If we utilised ${}^1\mathrm{H_2}$ as the loading gas, this would occur at 1~Da, and the majority of the signal observed there would be false hydrogen that did not originate from the sample - i.e.\ in the deuterated case, the peak at 2 is smaller than the peak at 1. It is thus clear that attempts to utilise the 1~Da peak will suffer severe problems with signal to noise, and must be treated with a commensurate level of uncertainty when attempting to interpret this peak, even in hydride formers such as Pd. This, however, strongly reinforces the need for deuterium as an isotopic tracer - experiments where this has not been done will likely  yield false results when interpreting hydrogen data. It is further likely that this effect will be more pronounced in materials with a reduced solubility, such as steels. Fully understanding these effects will likely require extensive modelling of the system under observation - it is known, for example, that the exact hydrogen concentrations observed are a function of the analysis parameters~\cite{Sundell2013}.

\begin{figure}[h]
 \centering
 \includegraphics[width=0.75 \textwidth]{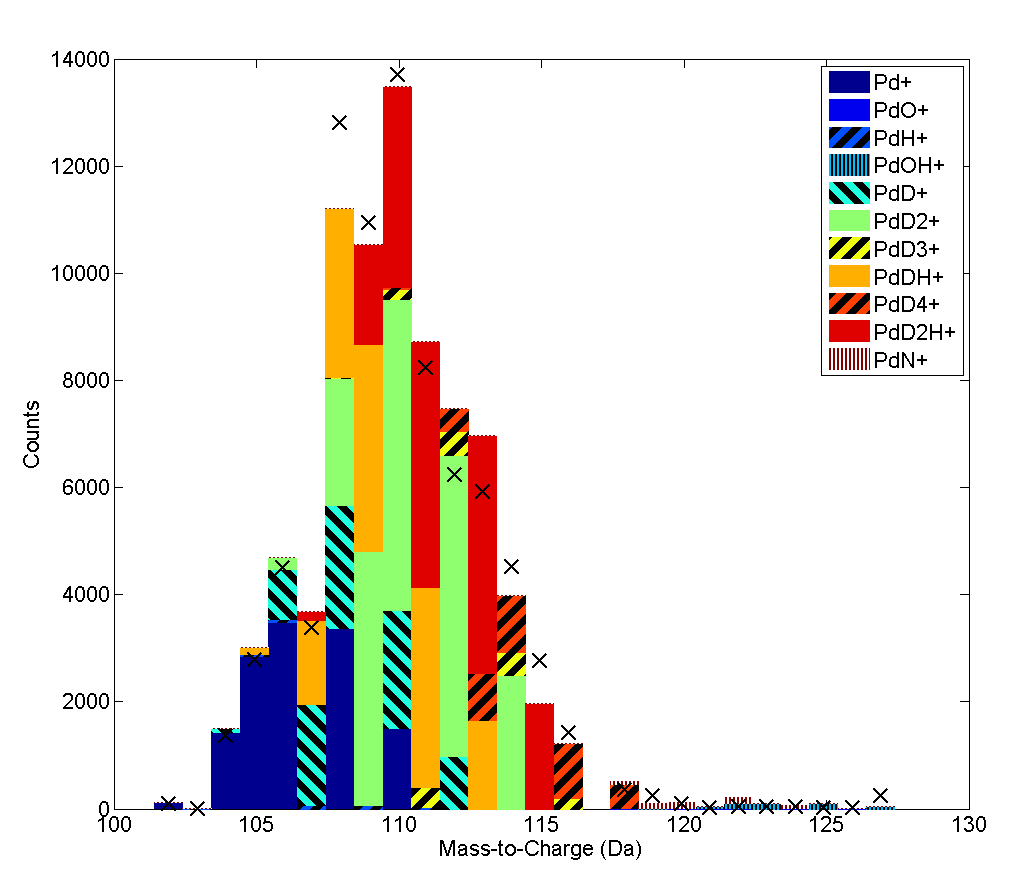}
 % PdD_H_peaks_decomposition.png: 0x0 pixel, 300dpi, 0.00x0.00 cm, bb=
 \caption{Decomposition of Pd$^+$/$\mathrm{PdH}^+_x$/$\mathrm{PdD}^+_x$ system (and additional peaks), here showing significant uptake of D in this peak. Markers are observed values, bars are estimated decomposed concentrations. The problem here can be solved using standard analysis approaches, as the overlap grouping has sufficient information available, such that the overlap is full rank, unlike the 1-6~amu grouping (Figure~\ref{fig:iterativeDecompositionH}). Note $\mathrm{PdH}^+_{2x}$ cannot be distinguished from $\mathrm{PdD}^+_x$, but is assumed to be small due to minimal $\mathrm{PdH}^+$.}
 \label{fig:pdHDecomposition}
\end{figure}

The Pd/PdH/PdD${}^+$ peak however, has sufficient rank to enable decomposition of this group (i.e.\ the problem is sufficiently constrained to provide a unique solution). The decomposition for this is shown in Figure~\ref{fig:pdHDecomposition}, where the observed spectrum is reasonably accurately reproduced, enabling quantitative extraction of the deuterium content of the peak. To accurately reproduce the observed peak profile, we consider PdOH${}^+$ and PdC${}^+$ formation, in addition to $\mathrm{PdD}_x^+$. There are limitations to this approach however, such as the $\mathrm{PdH}_{2x}^+$ and $\mathrm{PdD}_x$ isotopic fingerprints being effectively identical at detectable concentrations, and thus not able to be separated. However, as there is limited PdH and the pre-charge run did not show significant quantities of $\mathrm{PdH}_2$, it is considered that the effective abundance of $\mathrm{PdH}_2$ is small, and most H is present as PdOH.\@ Additionally there was a change in the charge state ratio, where prior to charging there is minimal $\mathrm{Pd}{}^{2+}$ present however after charging, the charge state ratio ($\mathrm{Pd}^{+}:\mathrm{Pd}^{2+}$) changes from 3 to 10. This value needs to be treated with caution, as there is an additional decrease in the $\mathrm{Pd}^{+}$ value, due to $\mathrm{PdD}^{x+}$ complex formation - however including the $\mathrm{PdD}^{+}$ contributions can only increase this ratio change beyond 10. It is thus clear that the field required to evaporate the sample has effectively decreased - presumably due to the presence of D adsorbates.

From these experiments it seems that the information obtainable from paired measurements (pre and post charging) are important for the estimation of overlapping species, as per the case in Figure~\ref{fig:iterativeDecompositionH}. Without such experiments - here paired on the same needle, but possibly over an ensemble of samples -the capability for quantitative analysis of spectra is diminished. This highlights the need for careful use of information from deuterated sources, and the comparison to uncharged and/or natural-H charged samples. Whilst some inaccuracies remain, other sources of error have not yet been investigated, such as detector response to multi-hit or closely timed hits~\cite{Rolander1994}, almost certainly occurring when mass separation is low.

\FloatBarrier

\subsection{Magnesium}

\begin{figure}[h]
 \centering
 \includegraphics[width=1.0 \textwidth]{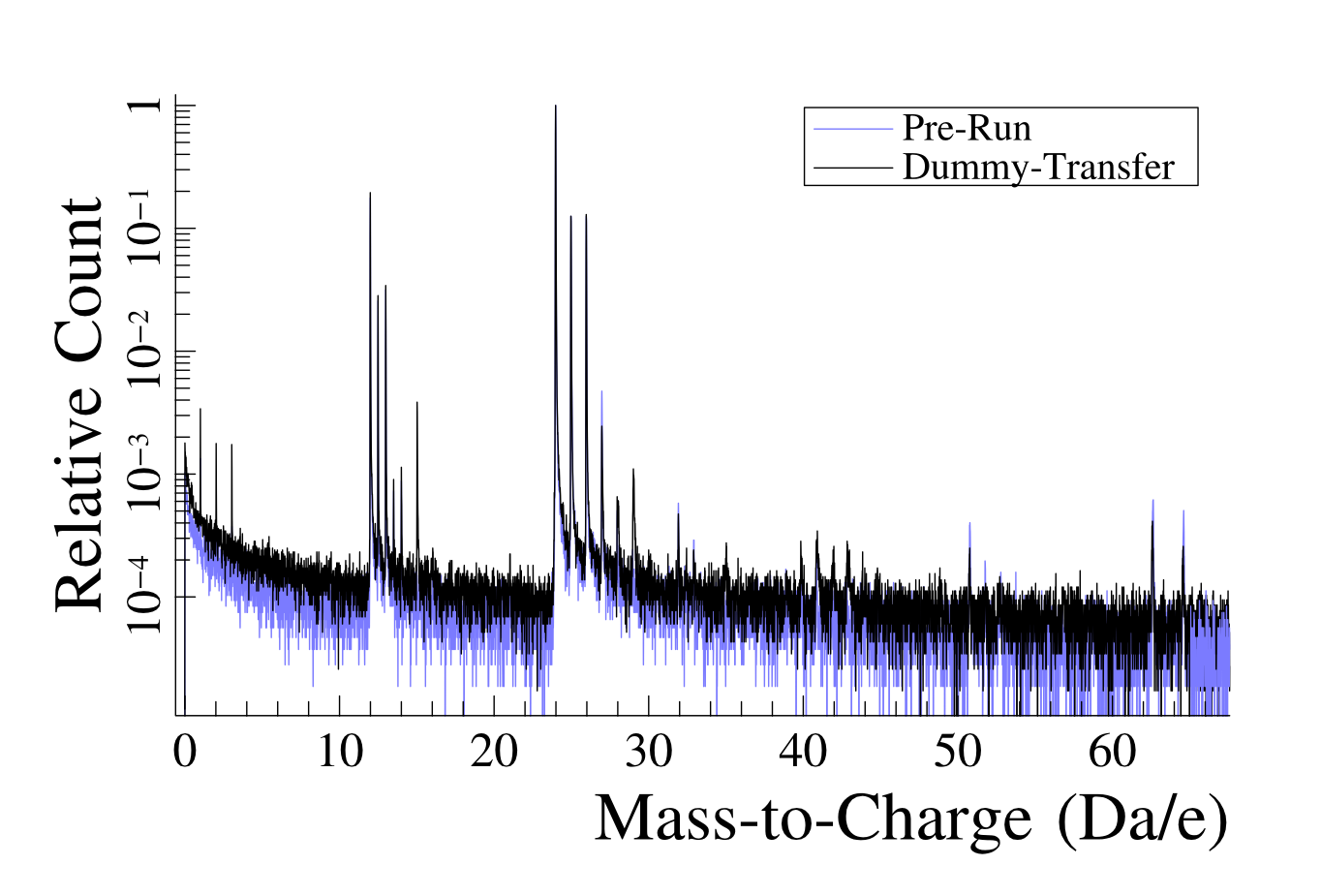}
 % PdD_H_peaks_decomposition.png: 0x0 pixel, 300dpi, 0.00x0.00 cm, bb=
 \caption{``Dummy'' run for Mg oxidation, showing the effect of sample retainment (36~hrs) under HV conditions as compared to the end of a prior atom probe run. Additional peaks are visible due to H, O , C and N, however Mg content is still high.}
 \label{fig:mgDummyRun}
\end{figure}

In an attempt to improve our understanding of Mg oxidation, samples were exposed to $\mathrm{O}_2$ and $\mathrm{H_2O}$. In initial experiments, the resulting oxide layers were very thin, and it was deemed necessary to check the effects of sample retainment within HV conditions, to ensure that contaminants could be distinguished from reaction products. Thus to identify contaminants that adsorb onto the tip during sample transfer, a dummy Mg sample was first field evaporated in the atom probe to produce a clean Mg surface. The sample was then transferred to the buffer chamber, where it remained for 36~h prior to being transferred  into the analysis chamber.  At the end of the initial cleaning run, primarily $\mathrm{Mg}^{+}$ and $\mathrm{Mg}^{2+}$ were detected as evaporating from the tip surface, as seen in Figure~\ref{fig:mgDummyRun}. 

Upon re-examination of this sample, after being placed within the `buffer' chamber of the atom probe ($\approx 1 \times 10^{-8}$~mBar) for 36~hrs, a number of contaminant species were observed, including H at 1-3~Da as well as peaks at 15~Da, most likely $\mathrm{NH}^{+}$, peaks at 27-29, 35 and 41-43~Da which are most likely combinations of H, O, C and N. After initial removal of this adsorbed layer, the Mg spectra returns to its original composition, with $\mathrm{Mg}^+$ and $\mathrm{Mg}^{2+}$. Due to the shallow absorption layer of the contaminant species, as compared to the oxides developed in our experiments, these peaks do not appreciably interfere with the identification of oxidised species in oxygen exposed runs.

%This is R14_24998 (Post hold) and R14_24993 (Pre-hold)

%This run is  R14_24436/R14_24453 (Mg40Fe) and Mg880Ge 
\begin{figure}[h]
 \centering
 \includegraphics[width=0.75 \textwidth]{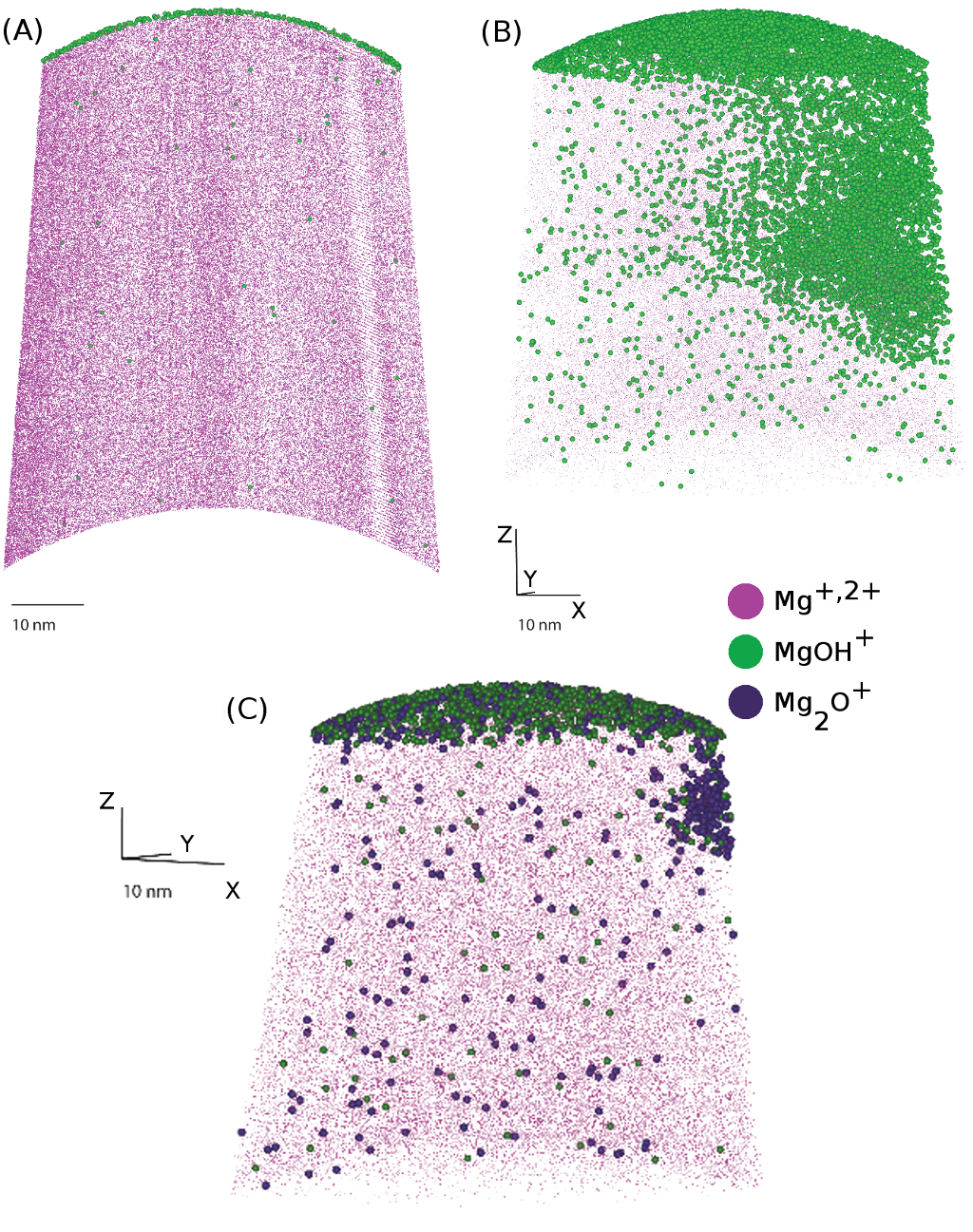}
 % PdD_H_peaks_decomposition.png: 0x0 pixel, 300dpi, 0.00x0.00 cm, bb=
 \caption{Mg samples Mg40Fe (a), Mg880Ge (b) and Mg650Fe (c) after oxidation at 2.5h @ 200~mBar, 1~h @ 500~mBar and 5~min @ 700~mBar respectively, all at room temperature. Oxidation front is clearly visible in each case, but a more extensive oxidation front is visible in the Mg880Ge and Mg650Fe samples.}
 \label{fig:mgSurfaceOxidation}
\end{figure}

After exposure to oxidising environments within the reaction cell, a clear oxide layer formed, however with two differing morphologies. In the instance of the Mg40Fe samples, O from the reaction cell adsorbed onto the surface of the samples, shown in Figure~\ref{fig:mgSurfaceOxidation}a. In the case of the sample with higher alloying content, the Mg880Ge  and Mg650Fe alloys appears to undergo a more rapid oxidation reaction (Figure~\ref{fig:mgSurfaceOxidation}b and c respectively).  The increased oxidation rate for these two samples is thought to be related to the higher quantity of solute elements. This is as Ge and Fe atoms on the surface can create a reactivity difference on the surface of the material, enhancing the difference in corrosion rate at differing positions on the surface~\cite{Thomas2015} -- within the samples here, Ge is $>$10 times more concentrated than Fe.

\section{Discussion}

In the case of the Pd alloy contamination, this likely did not originate from the stage or puck assembly, as the puck itself was fabricated from Al.  The uptake of hydrogen appears to have a significant surface component, which lies outside the bulk equilibrium values that are to be expected. When considering the solubility of Pd, and the enhancement factor from deformation (1.3) this yields a ratio of $1.3 \times 0.7 = 0.91$~\cite{Manchester2000}, which is very different from the observed $\approx 5.4:1$ D:metal ratio. This is in conjunction with the problem that it is impossible to fully separate the relative contributions of $\mathrm{PdH}_{2x}$ vs $\mathrm{PdD}_x$, without making some assumptions about the relative rates of formation of PdH and PdD. This alters the numerical problem from being a pure linear problem, to a constrained linear problem, where the relative intensities of two differing species must be kept in a fixed ratio. Further experiments are now required to fully ascertain surface effects, and the validity of these assumptions in various hydride forming systems.

Indeed, whilst atom probe examines the material at the near-atomic scale, the volumes of analysis are limited. To obtain a wider overview of the material under analysis, as well as to determine energetic information on hydrogen bonding, additional experiments such as deuteration in Secondary Ion Mass Spectrometry, or Thermal Desorption Spectroscopy will prove critical in fully understanding the nature of hydrogen within these materials.

However, the reaction cell configuration shown here is not only limited to hydrogen studies. We have additionally shown that oxidation research can be undertaken using this system, as highlighted in the study of the oxidation of Mg alloys under pure oxygen and air atmospheres. This work has shown that the species examined here are not artefacts of the transfer and vacuum storage process, but are real oxidation products. It is clear that the reaction cell configuration outlined here can successfully examine Mg samples, and will be effective in examining the development of oxides within these alloys.  Initial results from this new experimental method are promising in regard to developing our understanding of Mg corrosion processes and could provide significant progress towards expanding the use of lightweight Mg alloys in industries such as aerospace and biomedical technologies.

Thus this unique configuration allows for careful control of gas exposure and its effects on microstructures. This configuration will be critical in unravelling the effects of environmentally sensitive behaviours at the nanoscale. Thus through the use of pressure and temperature control, over the possible ranges allowed with this equipment, a greater understanding of the interactions of applied atmospheres can be obtained. This is an increasingly important area for atom probe tomography, particularly spurred by the need for greater understanding of oxidation, corrosion and hydrogen behaviours. Combining the observations here in the case of $\mathrm{H_2O}$ with further deuteration will yield significant new insights into the oxidation and hydrogenation mechanisms in light-alloy materials. 

\section{Future enhancements}
Further enhancements to the apparatus will enable higher temperature operation, and reduce contamination concerns during operation, such as the incorporation of Ni-based stage components to replace the temperature constraint arising from the use of either Cu or Al components. Further software enhancements will enable complex control of stage temperatures for specific thermal treatments, and enable the direct quantification of pressure-temperature-composition space at these length scales. 

Additional experiments to obtain temperature/pressure isotherms from this sample will assist in converting the experimental technique of hydrogen charging from a qualitative measurement to a fully quantitative methodology.

\section{Conclusions}

We have demonstrated the development of an in-situ reaction cell that can successfully perform controlled loading and unloading of gasses, using the exemplar of deuterium, oxygen and water in air gasses. We have shown that hydrogen can be detected, and through the use of modern decomposition software, some semi-quantitative analyses can be performed. Pairing or otherwise coupling data between deuterated and non-deuterated samples is critical to extracting the maximal information available in hydrogen analysis.  By demonstrating the effectiveness of this system for H charging using Pd, this apparatus provides a new way to investigate hydrogen quantitatively in materials with a high propensity to take-up hydrogen. Through the example of Mg, we have shown that we can investigate the effects of oxidation in differing alloys as a function of environmental exposure.

\section{Data access}
Data for this publication is available on the Oxford Research Archive, http://ora.ox.ac.uk

\section{Acknowledgements}

D.\ Haley, M.\ P.\ Moody and P.\ A.\ J.\ Bagot wish to acknowledge the EPSRC HEMS project EP/L014742/1. I. McCarroll and J. Cairney acknowledge the facilities, and the scientific and technical assistance, of the Australian Centre for Microscopy and Microanalysis Research Facility at The University of Sydney.  The authors would also like to acknowledge Ruiliang Liu, Sebastian Thomson and Nick Birbilis (Monash University) for kindly providing the Mg-Ge alloys used in this work.

\bibliographystyle{unsrt}
\bibliography{rxncell.bib}

\end{document}